\documentclass{aa}
\usepackage{graphicx}
\usepackage{natbib}
\usepackage{rotating}
\usepackage{txfonts}

\bibpunct{(}{)}{;}{a}{}{,}
\def\aap{A\&A}

\def\aj{AJ}
\def\apj{ApJ}
\def\apjs{ApJS}
\def\apss{Ap\&SS}
\def\apjl{ApJ}

\def\mnras{MNRAS}

\def\pasj{PASJ}

\def\aapr{A\&ARev}

\def\jrasc{JRASC}

\def\apph{Astropart. Physics}

\begin{document}

\title{An investigation into the fraction of particle accelerators among colliding-wind binaries} 
\subtitle{Towards an extension of the catalogue}

\author{M. De Becker\inst{1}, P. Benaglia\inst{2,3}, G.~E. Romero\inst{2,3}, and C.~S. Peri\inst{2,3}}

\offprints{M. De Becker}

\institute{Space sciences, Technologies and Astrophysics Research unit -- STAR, University of Li\`ege, Quartier Agora, 19c, All\'ee du 6 Ao\^ut, B5c, B-4000 Sart Tilman, Belgium 
\and
Instituto Argentino de Radioastronom\'{\i}a, CCT-La Plata, CONICET, Argentina
\and
Facultad de Ciencias Astron\'{o}micas y Geof\'{\i}sicas, Universidad Nacional de La Plata, Paseo del Bosque s/n, 1900 La Plata, Argentina
\\
}

\date{Received ; accepted }

\abstract
{Particle-accelerating colliding-wind binaries (PACWBs) are multiple systems made of early-type stars able to accelerate particles up to relativistic velocities. The relativistic particles can interact with different fields (magnetic or radiation) in the colliding-wind region and produce non-thermal emission. In many cases, non-thermal synchrotron radiation might be observable and thus constitute an indicator of the existence of a relativistic particle population in these multiple systems. To date, the catalogue of PACWBs includes about 40 objects spread over many stellar types and evolutionary stages, with no clear trend pointing to privileged subclasses of objects likely to accelerate particles. This paper aims at discussing critically some criteria for selecting new candidates among massive binaries. The subsequent search for non-thermal radiation in these objects is expected to lead to new detections of particle accelerators. On the basis of this discussion, some broad ideas for observation strategies are formulated. At this stage of the investigation of PACWBs, there is no clear reason to consider particle acceleration in massive binaries as an anomaly or even as a rare phenomenon. We therefore consider that several PACWBs will be detected in the forthcoming years, essentially using sensitive radio interferometers which are capable of measuring synchrotron emission from colliding-wind binaries. Prospects for  high-energy detections are also briefly addressed.}


\authorrunning{Authors}
\titlerunning{Fraction of PACWBs among CWBs}

\maketitle

\section{Introduction}\label{intro}

Galactic objects able to accelerate particles up to relativistic energies are numerous. Though most Galactic cosmic rays are thought to be accelerated in supernova remnants \citep{Decourchelle,VinkSNRutrecht}, other categories of stellar objects are likely to act as particle accelerators as well, among others $\gamma$-ray binaries \citep{valenti2011,dubus2013}, microquasars \citep{valenti2007}, bow-shock runaways \citep{ntbowshock,delValle2015}, or particle-accelerating colliding-wind binaries \citep{catapacwb}. 

Colliding-wind binaries (CWB) consist of systems made of at least  two massive stars, with OB-type or Wolf-Rayet (WR) classifications, whose stellar winds collide. Particle-accelerating colliding-wind binaries (PACWBs) constitute a subset of CWBs presenting strong indication that a particle acceleration process is at work. Indicators for particle acceleration are non-thermal emission processes involving relativistic electrons (synchrotron radiation, inverse Compton scattering) and/or relativistic protons (such as neutral pion decay). In the so-called standard scenario for PACWBs, the wind collisions lead to strong shocks in the wind-wind interaction regions, where diffusive shock acceleration (DSA) can operate to produce a population of relativistic particles. Though other acceleration mechanisms such as magnetic reconnection may be invoked in some astrophysical environments  \citep[e.g.][]{lazarian2012,mhdcwbradio2012}, DSA is thought to be the best working hypothesis to explain the presence of non-thermal particles in these objects \citep[e.g.][]{EU,BRwr,PD140art,Pit2,RPR2006,debeckerreview,reitberger2014}.

The catalogue of PACWBs currently includes 43 objects \citep{catapacwb}. The diversity of stellar types in the catalogue does not point to any spectral classification likely to favour particle acceleration. To date only a fraction of colliding-wind binaries are known to be PACWBs (i.e. having detected non-thermal emission) and we do not know yet what is the fraction of this sample explained by observational biases or caused by actual physical differences between particle accelerators and other massive binaries. If the actual census of PACWBs is mostly due to observational biases, it would suggest that the whole category of massive binaries may contribute to the production of soft Galactic cosmic rays. However, if the actual census is not only explained by observational biases, it would mean we missed something really important in the physics that triggers particle acceleration in these objects. 

The fraction of PACWBs among colliding-wind binaries and its determination are highly important for several reasons, especially if the fraction is not small. First of all, taking into account particle acceleration and non-thermal processes could be important  for the modelling of the physics of massive binaries. On the other hand, the whole population of PACWBs may be significant for the production of low-energy Galactic cosmic rays. Colliding-wind binaries are indeed able to accelerate particles for the whole duration of their pre-supernova evolution phase. This allows for the possibility that the integrated contributions -- over the evolution timescale --  from all these systems may be significant. For instance, with an energy injection rate in charged particles between 0.01\,$\%$ and 1\,$\%$ \citep[see e.g.][]{catapacwb}, a massive binary will, on average, convert 10$^{32}$--10$^{34}$\,erg\,s$^{-1}$ into relativistic particles. Assuming a Galactic population of about 10$^5$ massive stars, the energy production rate of cosmic rays by pre-supernova massive stars could be about 10$^{37}$--10$^{39}$\,erg\,s$^{-1}$. This is significant  compared to the total power in cosmic rays of the order of 10$^{40}$\,erg\,s$^{-1}$ \citep{vinksnr2004}, especially in the lower energy part of the cosmic-ray spectrum where PACWBs can contribute (however, a significant contribution above a fraction of TeV is unlikely because of severe cooling losses in the acceleration region; \citealt[][]{debeckerreview,reitberger2014}). So, depending on the fraction of particle accelerators among massive stars, their total contribution to the production of Galactic cosmic rays could be important.

The determination of such a fraction requires two elements: (i) a good census of particle accelerators (e.g. notably by identifying new radio synchrotron emitters or high-energy non-thermal emitters) and (ii) a good census of binaries among massive stars. However, both censuses are affected by strong observational biases, resulting in a very poor determination of the PACWB/CWB fraction. The present number of PACWBs is undoubtedly a lower limit because of the difficulty in identifying clearly non-thermal emission processes used as indicators for particle acceleration. For instance, synchrotron radio emission may be significantly absorbed by the stellar wind material in a large fraction of the orbit, or inverse Compton X-ray emission may be too weak or overwhelmed by thermal emission \citep[see][and references therein for discussions about these issues]{catapacwb}, thus preventing us from identifying their particle accelerator status. On the other hand, the binary fraction among massive binaries is also a controversial issue.  It seems that a huge fraction of  O-type stars are found in binary or higher multiplicity systems \citep[see e.g.][]{sana2012,sana2014}, even though a clear determination of this binary fraction requires the extensive use of many complementary observational techniques and it is still a matter of debate.

The goal of this paper is to address critically the question of the fraction of particle accelerators among massive binaries. More specifically, it intends to explore the current knowledge on massive stars to derive valuable tracks for the selection of candidates to be investigated with the aim of identifying new particle accelerators. The goal is also to present guidelines for adequate observational strategies. The paper is organized as follows. Section\,\ref{extension} discusses several criteria worth exploring for the selection of new candidates among known stellar populations. Adequate observation strategies, based on present and future observatories, are then briefly discussed. The summary and conclusions are given in Sect.\,\ref{concl}. 

\section{Towards an extension of the catalogue}\label{extension}
In order to improve our estimation of the PACWB/CWB fraction, a crucial requirement is the in-depth investigation of the category of PACWBs itself. This translates into the necessity of investigating other colliding-wind binaries to find new members to include in the catalogue. To do so, it seems  reasonable first to establish tentative criteria for selecting candidates to be observed notably in the radio domain to search for the presence of synchrotron emission, which is the main tracer of particle acceleration in massive binaries. These criteria deserve to be discussed taking into account recent relevant information (see Sect.\,\ref{selection}) to establish a list of potential candidates. Second, some guidelines for observation strategies (see Sect.\,\ref{strategies}) need to be established to explore the sample of candidates and potentially identify new members of the catalogue. We stress here that the criteria discussed in Sect.\,\ref{selection} are by no means strict indicators of particle acceleration, but are rather criteria for making a pre-selection of massive binaries worth investigating in order  to subsequently search for non-thermal radiation revealing the action of a particle acceleration process.

\subsection{Selection criteria}\label{selection}

\subsubsection{Stellar wind parameter space}\label{wind}

It is certainly worth asking whether PACWBs occupy a well-defined area of the stellar wind parameter space, in particular in terms of mass loss rate and terminal velocity. These two parameters combined  determine the kinetic power of a given stellar wind. Because the injection of kinetic power into the wind-wind interaction region constitutes the source of energy driving the particle acceleration mechanism, one may indeed expect the most powerful stellar winds to favour the detection of their associated non-thermal radiation. A discussion on the energy budget in PACWBs is developed in \citet{catapacwb}.

Considering the spectral classifications of all members of the catalogue, we find that PACWBs span the complete range of O-type and WR-type stars\footnote{By convention, O-type and WR-type members of the catalogue are objects whose primary (or more evolved) component is of spectral type O and WR, respectively. For instance, a WR + O system is considered to be a WR-type binary, even  though one the components belongs to the O-type category. Instead, a WR + B and a O + B system are not  considered  B-type systems, but  WR- and O-type systems, respectively.}. There is, however, a gap in the distribution of primary stellar wind parameters (see Fig.\,1 in \citealt{catapacwb}) corresponding to the weaker winds of B-type stars. This gap comes from the lack of PACWBs identified so far with the more massive or more evolved component belonging to the category of B-stars. These later-type objects are characterized by slower stellar winds and lower mass loss rates than O-type stars. The only exception is HD\,190603, which is an early B star with enhanced mass loss rate explained by its evolutionary stage, probably on the way to some kind of luminous blue variable phase \citep{clarkBHG}. 

\begin{figure}[ht]
\begin{center}
\includegraphics[width=80mm]{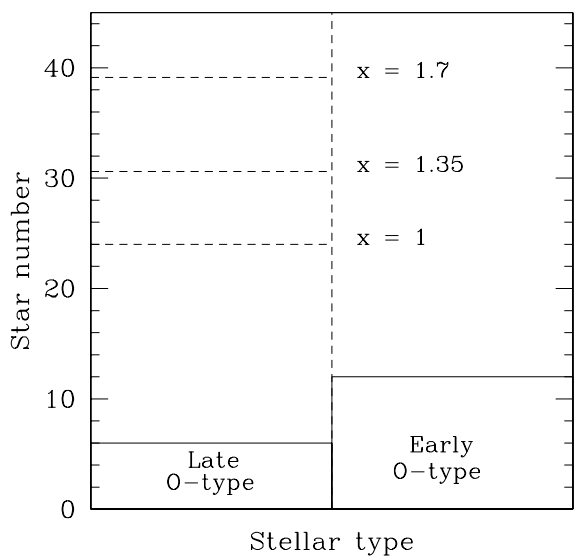}
\caption{Expected numbers of late-type stars (left column, horizontal dashed lines) given the number of early-type stars (right column) in the catalogue of PACWBs for different assumptions on the IMF index x. The solid horizontal lines stand for the numbers found in the current catalogue of PACWBs.\label{imf}}
\end{center}
\end{figure}

In addition,  the catalogue is very poor in late-type O-systems, especially on the main sequence, even though these types are expected to be much more abundant than earlier objects. In order to roughly quantify this apparent deficit in late-type objects in the catalogue with respect to the overall massive star population, it is possible to compare the O-type content of the catalogue to expectations based on standard initial mass functions (IMF). The inclusion of WR stars in this discussion is not straightforward considering the uncertainty on the initial masses of these evolved objects likely to  significantly blur the results. According to \citet{imfschaerer}, resolved stellar populations show a reasonable agreement with the classical Salpeter IMF \citep{salpeter}, even for stellar masses higher than 10\,M$_\odot$. On the basis of this idea,  the numbers given by \citet{zinneckerreview} can be used for stars in well-defined mass ranges. These authors consider the following categories: late O-type stars (between 16 and 32\,M$_\odot$), early O-type stars (between 32 and 64\,M$_\odot$), and O/WR-type massive stars (between 64 and 128\,M$_\odot$). We focus on the late and early O-type categories. Reasonable masses taken from \citet{martins} can be applied to all O-type members of the catalogue (i.e. O + O and O + B systems) to include them in the adequate category. This leads to six late O-type and 12 early O-type systems actually present in the catalogue. Assuming different values for the index of the IMF, \citet{zinneckerreview} anticipate star numbers clearly indicating that late-type objects dominate the stellar population. The situation is summarized in Fig.\,\ref{imf}, assuming an IMF of the form $\mathrm{dN/d\log\,M \propto M^{-x}}$, where the number of objects is presented in the form of a two-element histogram, one for early-type objects (right) and one for late-type objects (left). The horizontal solid lines stand for the actual numbers (respectively 6 and 12) found in the catalogue for early- and late-type systems, respectively. Given the uncertainty on the IMF in the high-mass range, we decided not to restrict our discussion to one value for x and we adopted three values \citep{zinneckerreview}: x = 1, 1.35 (Salpeter IMF), and 1.7. For the six early O-type systems found in the catalogue, those IMF distributions allow us to predict numbers of expected late-type systems represented by the horizontal dashed lines in the left part of Fig.\,\ref{imf}. It is clear from this plot that late O-stars actually in the catalogue are heavily under-represented. In other words, more early O-type objects than expected populate the catalogue of PACWBs with respect to a classical overall stellar population. For instance, assuming a Salpeter IMF index of 1.35 and considering the six early O-type systems in the catalogue, the actual number of late O-stars is a factor 5 lower than expected assuming a standard stellar population.

The lack of detection of B-type PACWB, along with the relative weak number of late-type main-sequence O-stars in the catalogue, could potentially be explained by two main factors. First, one might think that stars with weaker stellar winds are not able to accelerate particles. This scenario seems unlikely at first sight  as there is a priori no reason to turn off the DSA mechanism below a given threshold of kinetic power. As counter examples, we mention the cases of particle acceleration in the heliosphere or at the Earth's bow shock \citep{lee2012,burgess2012} where the kinetic power is much lower than for PACWBs though the acceleration mechanism is still operating. Certainly less energy would be available to drive the shocks, but it does not necessarily mean that the acceleration mechanism would not be operating. On the other hand, stars with weaker winds may be able to accelerate particles, but the associated radiative process may be operating at an emission level below the sensitivity of past and present observations. It is not possible to  state that PACWBs must be made exclusively of O- and WR-type stars, but  the present ability to detect a non-thermal signal is probably not enough to warrant detection of the weakest emitters, i.e. those powered by weaker stellar winds. As a result, the favoured classes of objects likely to give rise to new identifications of PACWBs in the coming years are O stars (of any type, but with a lower priority on late-type main-sequence stars) and WR stars (including WN and WC objects of all subtypes).

\subsubsection{Orbital period}\label{period}

To date, only a fraction of the catalogue members have well-determined orbital periods. Figure\,2 in \citet{catapacwb} clearly shows that the range of orbital periods goes from a few weeks  to several decades. The shortest period system with well-determined orbital elements is Cyg\,OB2\,\#8A (\object{Schulte 8}, O6If + O5.5III(f)). This system presents a clear modulation of the radio light curve with a period of about 22 days \citep{Blomme8apaper}, in agreement with the orbital solution derived from spectroscopic data \citep{Let8a}. The catalogue also includes  Cyg\,OB2-335 (\object{MT91 771}) with a period of about 2 days \citep{kiminkicygob2} and with very clear indications of synchrotron radio emission \citep{setia}. However, this object should be investigated to search for the presence of a potential third object whose wind interaction with the close binary could be a more appropriate site for the production of a detectable synchrotron radio emission (see \citealt{catapacwb} for a discussion). The reason is that it is unlikely that synchrotron radio emission from such a close system could escape from the optically thick winds of the close pair, but would rather emerge from a more open wind interaction region in a wider orbit. In addition, the efficient cooling of electrons due to inverse Compton losses may significantly reduce the population of relativistic electrons likely to contribute to the synchrotron emission. Aside from close binaries, the PACWB with the longest determined period is \object{15\,Mon}, with an orbital period of about 25 years \citep{gies15mon}. In addition, we  also note the cases of even wider systems, such as \object{HD\,93129A} \citep{benaglia2015} for O-type objects, or WR-type systems such as \object{WR\,146} \citep{wr146companion} and \object{WR\,147} \citep{williamswr147} whose periods are not determined but are certainly longer than a century\footnote{In the case of these two  systems, direct evidence for an orbital motion is still lacking, though the wind-wind interaction in a gravitationally bound binary system in those objects is the most likely scenario.}.

These examples clearly suggest that except for close binaries with periods of a few days, any orbital period may be adequate. There could be two reasons for the lack of PACWBs with periods of a few days. First of all, as the main indicator of particle acceleration is synchrotron radio emission, free-free absorption of radio photons by the stellar wind material is expected to be higher in short-period systems characterized by short stellar separations. This is likely to introduce some significant bias in the identification of PACWBs. Second, it is still not so straightforward that close binaries are able to significantly accelerate particles through the DSA mechanism. In addition to the likely inhibition of particle acceleration by the expected cooling caused by inverse Compton scattering of photospheric photons\footnote{The source might, nevertheless, accelerate protons efficiently.}, it has not yet been confirmed that the turbulent nature of shocked winds in short-period systems would easily allow particles to be affected by the shock velocity jump responsible for the acceleration (although turbulent magnetic reconnection remains a possibility in such contexts, \citealt{lazarian2012}). As a consequence, the most favoured massive binary systems likely to be investigated to identify new PACWBs are systems with periods longer than a few weeks. 

\subsubsection{Thermal X-rays from colliding-winds}\label{xrays}

In the so-called standard scenario for PACWBs, particle acceleration takes place in the wind-wind interaction region. The supply of energy driving the particle acceleration mechanism comes from the injection of kinetic power in this interaction region. In the context of colliding-wind binaries, it is known that a fraction of that kinetic power is converted into thermal X-rays, due to the heating of shocked plasma downstream of the strong shocks. As a result, post-shock temperatures can reach values of several 10$^7$\,K, and produce thermal X-ray spectra over the whole soft X-ray domain, below 10\,keV \citep[see e.g.][]{SBP,PS,pitpar2010}. A bright thermal X-ray spectrum produced by a wind-wind interaction region can therefore be considered as the signature of an efficient power injection into that region. One may thus wonder whether such a bright thermal X-ray spectrum could be a good indicator of a high-energy injection rate favourable to particle acceleration in that region.

To check for the validity of this idea, the case of some PACWBs already investigated in X-rays deserves to be examined. Some systems, such as \object{WR\,140} \citep{wil140,pollock140} or Cyg\,OB2\,\#8A \citep{harnden,DeBcyg8a}, are indeed known to be both well-studied PACWBs and bright thermal X-ray emitters, due to their colliding winds. However, other examples should be commented. The first  is HD\,167971. It consists of a hierarchical triple system with a 3.3-day O-type binary and a third later-type companion evolving on a 21-year orbit \citep{vlti167971,ibanoglu2013}. Recently, it has been demonstrated that its thermal X-ray spectrum is dominated by the colliding-wind region in the close binary, with only a weak/moderate contribution coming from the wind-wind interaction in the wide orbit \citep{debecker6604new}. However, it has also been demonstrated that \object{HD\,167971} presents a non-thermal radio emission modulated with a period of 21 years, thus providing  compelling evidence for a colliding-wind region origin \citep{Blo167971}. In addition, HD\,167971 is the brightest O-type synchrotron radio emitter included to date in the catalogue of PACWB. Another relevant example is  HD\,93129A. It is known as an early O-type wide binary with undetermined period (probably more than a century), with non-thermal radio emission recently imaged in Long Baseline Array observations \citep{benaglia2015}. The direct imaging of this emission region points to a clear and significant synchrotron radio emission coincident with the colliding-wind region. However, the investigation in X-rays by \citet{carinachandra} did not point to any strong over-luminosity attributable to an X-ray bright wind-wind interaction region. These examples demonstrate that the association between bright thermal X-ray emission and bright non-thermal radio emission is not necessarily clear. As a result, the quest for new members of the catalogue should by no means be restricted to objects known to display a bright thermal soft X-ray spectrum.

\subsubsection{Surface magnetic field strength}\label{magn}

The PACWB status is mostly attributed on the basis of the detection of synchrotron radio emission. One may thus wonder whether a significant -- or even strong -- magnetic field is needed to identify a PACWB. So far, only a few PACWBs have been investigated using spectropolarimetric techniques aimed at measuring surface magnetic fields. In particular, \citet{neiner2015} applied such techniques to a sample of nine O-type PACWBs in the southern hemisphere. The measurements did not lead to any detection of magnetic fields at the level of a few tens of gauss or higher. 

These results lend support to the idea that strong surface magnetic fields do not constitute a requirement to belong to the category of PACWBs. This is not really surprising considering the expected strength of magnetic field needed to explain the measured synchrotron radio emission in some massive binaries. For instance, the in-depth investigation performed by \citet{Doug} points to an order of magnitude of a few mG for the local magnetic field strength (assuming equipartition), i.e. in the synchrotron emission region. Provided this local magnetic field is of stellar origin, its strength results first from the geometrical dilution of the field intensity depending on the distance to the stellar surface. On the basis of the work by \citet{WD1967}, \citet{UM} described a few dependencies of the magnetic field as a function of the distance $r$. Although the dipole dependency is suitable for locations very close to the stellar surface, intermediate and longer distances call upon radial and toroidal regimes, respectively. For typical distances between the wind-wind interaction region and stellar surfaces in the sample of known PACWBs, the radial regime should apply

$$ B\,\propto\,B_{\rm s}\,\times\,\frac{R^2}{r^2},$$
\noindent where $B$ is the local magnetic field strength (at the stagnation point of the colliding-winds, assumed to be representative of the position of the synchrotron emission region), $B_{\rm s}$ is the surface magnetic field strength, $R$ is the stellar radius, and r is the distance from the star to the stagnation point.

With distances between the wind-wind interaction region and the stellar surface ranging between 50 and 4000\,R$_\odot$ (depending on the orbit and on the orbital phase), the geometrical dilution alone would convert a local field strength of 1 mG to values between 0.025 and 50 G at the surface (assuming a stellar radius of 20\,R$_\odot$). This range is of the order of  or even fainter than  the typical sensitivity of current facilities used to measure stellar magnetic fields. If  a likely magnetic amplification factor is now considered in the wind-wind interaction region (see e.g. \citealt{mhdcwbradio2012}), the typical value of 1\,mG should be reduced by the amplification factor before applying the geometrical dilution factor. Depending on the regime of the shocked wind -- adiabatic or radiative -- the structure of the shock is expected to be quite different. An adiabatic shocked wind will amplify the local magnetic field mostly close to the contact discontinuity. However, a radiative shocked wind can give rise to a magnetic amplification in the wind-wind interaction region 2 or 3 orders of magnitude larger than in the adiabatic case, with a more complex structure due to its turbulent nature \citep{mhdcwbradio2012}. Magnetic amplification is highly important as it will further reduce the expected and sufficient value for the typical surface magnetic field strength required for the production of synchrotron radiation at level of current radio detections. Furthermore, a potential amplification of the ambient magnetic field by accelerated charged particles themselves cannot be ruled out, as suggested in the case of supernova remnants \citep{lucekbell2000}. These facts clearly show that the search for new PACWBs should by no means be restricted to strongly magnetic objects.

\subsection{Observation strategies}\label{strategies}

\subsubsection{Radio observations}\label{radio}
The main tracer of particle acceleration in massive binaries is synchrotron radio emission; hence, radio astronomy constitutes a privileged tool to identify new members of the catalogue. However, several facts must be taken into account at the hour of planning radio observations. 

The radio spectrum of PACWBs is a combination of thermal (free-free) and non-thermal (synchrotron) emission. The thermal part is expected to be stationary over the typical evolution timescale as it is related to the stellar wind properties. However, the synchrotron component should vary as a function of the orbital phase for two reasons. First, the synchrotron emission will depend on the stellar separation in the system, especially because of its incidence on the local magnetic field strength (provided the magnetic field is of stellar origin and not produced in the wind-wind interaction region). The emitted non-thermal component is expected to brighten towards periastron. Second, free-free absorption will significantly reduce the measured synchrotron component with an amplitude that depends on the orientation and geometry of the system. As a result of these two factors, synchrotron radiation will not necessarily be easy to detect at any phase of the orbit. This motivates therefore multiple observations likely to reveal a variability in the radio emission, legitimately considered as an indicator of non-thermal emission in the wind-wind interaction region. In summary, observation strategies for massive binary systems should consider measurements at several epochs to search for a significant variability in the radio emission. The privileged orbital phases – provided the orbital elements are known – should be close to periastron and apastron to optimize the expected variation. It should be remembered,  however, that the variation of the synchrotron component can be severely modulated throughout the orbital motion by a number of mechanisms, including free-free absorption, in addition to changing radiation and magnetic field density. Consequently the maximum and minimum of the radio flux density will not necessarily coincide with those extreme orbital phases.

Another issue concerns the observation frequencies. Synchrotron emission in the optically thin regime decreases as a function of increasing radio frequency, but free-free emission increases with increasing frequency. These two antagonist behaviours allow us to identify whether the radio emission (in the investigated spectral range) is thermal or non-thermal on the basis of the spectral index. This requires radio measurements at at least two frequencies. In addition, decimeter wavelengths should not be neglected as synchrotron radiation is expected to dominate free-free emission in that region of the spectrum. However, the strong impact of free-free absorption at longer wavelength can constitute a severe limitation at least for the shortest period systems (i.e. characterized by the shortest separations). Also, (quasi-)simultaneous observations at different wavelengths are crucial to determining spectral indices and unambiguously identifying  synchrotron emission. Detection of linear polarization also might be crucial in the identification. We note, however, that several depolarization mechanisms can operate in the complex environment of the colliding wind region \citep{pola146}. This and the small fluxes implied make the detection of polarization an extremely difficult task.

Although the brightest PACWBs emit at the mJy level \citep[see][ and references therein for typical flux densities of the catalogue members]{catapacwb}, sensitivities below that level are needed to measure flux densities of the synchrotron emission region in many systems. The lack of detection of non-thermal radio emission for several massive binaries is certainly due in part to a sensitivity-related observational bias. This motivates a search for synchrotron radio emission at a lower level, even as low as the $\mu$J level, which is not achievable with current facilities.

Observation strategies should also include considerations about the angular resolution. High-resolution radio observations (subarcsec) with modern interferometers help to avoid source confusion in regions where most massive stars are located (generally in clusters or OB associations with numerous stellar sources). In addition, even better angular resolutions (a few milliarcsec) are needed in order to resolve the synchrotron emission region (coincident with the colliding winds) and the thermal emission from the individual stellar winds of the components of the system. The colliding winds of very few systems have been mapped in radio to date: WR\,140 \citep{Doug140}, WR\,146 \citep{oconnorwr146art}, WR\,147 \citep{williamswr147}, Cyg\,OB2\,\#5 \citep{contr,ortizcyg5}, Cyg\,OB2\,\#9 \citep{vlbi2006}, and HD\,93129A \citep{benaglia2015}. The northern e-MERLIN\footnote{http://www.e-merlin.ac.uk/}, the Very Long Baseline Array\footnote{https://science.nrao.edu/facilities/vlba} (targets with declinations greater than --40 deg), and the southern Australian Long Baseline Array\footnote{http://www.atnf.csiro.au/vlbi/} (targets with negative declinations) provide the appropriate configurations to observe and resolve PACWBs. The lowest observable bands are centred at 1.3, 0.5, and 2.3 GHz, respectively.

On the other hand, the facilities available to monitor the integrated flux of the stellar systems, i.e. with unresolved synchrotron emission region and individual stellar winds, with the appropriate sensitivity are the Jansky Very Large Array\footnote{https://science.nrao.edu/facilities/vla} (JVLA), the Australia Telescope Compact Array\footnote{http://www.narrabri.atnf.csiro.au/} (ATCA), and the Giant Metrewave Radio Telescope\footnote{http://www.gmrt.ncra.tifr.res.in/} (GMRT). The JVLA observes from about 70 MHz to 50 GHz, the ATCA from 1.1 to 105 GHz, and the GMRT from 50 to 1500 MHz. MeerKAT \footnote{https://www.ska.ac.za/science-engineering/meerkat/}, the precursor of the future Square Kilometer Array, will be made of 64 dishes in its final configuration (end of 2017). It will provide an angular resolution down to 0.8 arcsec at 2 cm and observing bands between 0.58 and 14.5 GHz \citep{meerkat}. With its expected specifications, MeerKAT will undoubtedly constitute a highly valuable tool for investigating PACWBs.\\

The census of radio measurements provided by \citet{catapacwb}  characterizes some kind of truncated banana-shaped region in the two-dimensional space defined as (i) the kinetic power ($P_\mathrm{kin}$) of the main stellar wind, and (ii) the radio synchrotron efficiency parameter (RSE) defined as the ratio between the radio luminosity and the wind kinetic power (see Fig.\,3 in \citealt{catapacwb}). The radio synchrotron efficiency parameter is thus equivalent to the fraction of the kinetic power that is converted into radio emission downstream of a series of energy conversion processes (see Fig.\,2 in \citealt{catapacwb}). A qualitative description of the adequate region can be achieved through a quadratic regression (in log scale) of its upper and lower boundaries. Data points from four objects located as close as possible to the apparent boundaries were used to characterize these curves. These systems are Cyg\,OB2-335, HD\,167971, WR\,112, and WR\,140 for the upper boundary and 15\,Mon, HD\,151804, WR\,21a, and WR\,133 for the lower boundary (see \citealt{catapacwb}). The procedure yields
\begin{equation}\label{up}
\log\,\mathrm{(RSE)}_\mathrm{Upper} = -287.0 + 15.94\,\log\,(P_\mathrm{kin}) - 0.226\,\big[\log\,(P_\mathrm{kin})\big]^2 
\end{equation}
\begin{equation}\label{low}
\log\,\mathrm{(RSE)}_\mathrm{Lower} = -357.5 + 20.09\,\log\,(P_\mathrm{kin}) - 0.2878\,\big[\log\,(P_\mathrm{kin})\big]^2 
\end{equation}
\noindent with correlation coefficients larger than 0.99 for both relations. 
\begin{figure}[ht]
\begin{center}
\includegraphics[width=9.0cm]{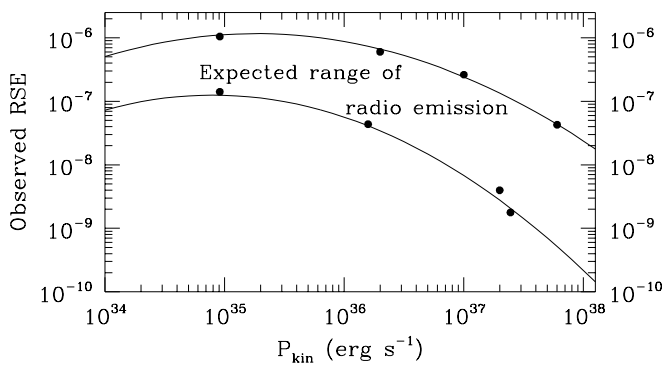}
\caption{Rough estimate of the expected region of radio emission as a function of the kinetic power based on the radio synchrotron efficiency (RSE) derived for the members of the catalogue of PACWBs ({\it truncated banana diagram}). The two curves correspond to Eqs.\,\ref{up} and \ref{low}. The black bullets represent the data points used for the quadratic regression. \label{tbd}}
\end{center}
\end{figure}

In principle, the plot shown in Fig.\,\ref{tbd}, for a given system of known spectral classification, allows the range of expected radio luminosities to be inferred on the basis of the kinetic power of the system. Knowing the spectral classification, typical values for the terminal velocity and the mass loss rate can be used to estimate the kinetic power. The expected lower and upper boundaries on the RSE can be computed on the basis of Eq.\,\ref{low} and Eq.\,\ref{up}, respectively. The radio luminosities can then be converted into a flux density range, assuming a spectral shape (i.e. spectral index) and an adequate distance. This procedure  provides a first rough guess of the expected range of flux density for a system intended to be investigated on the basis of the current census of measurements made on PACWBs with present radio facilities. We caution, however, that the above relations are based on the actual measurements made so far on PACWBs, and are therefore affected by the specifications of present radio observatories. This  approach does not  take into account the expected gain in sensitivity that could be achieved with future facilities.\\

Considering the growing importance of surveys conducted by various radio observatories, one may ask whether those surveys are of importance for the identification of new PACWBs. Taking into account the requirement that measurements should be obtained at various epochs (because of orbital modulation) and at various frequencies (to determine spectral indices), it is clear that surveys are not enough, justifying the  dedicated observations for carefully selected objects (on the basis of the criteria discussed in Sect.\,\ref{selection}). However, surveys can provide relevant information to prepare the dedicated observations; in particular, they will allow  emission levels or upper limits to be established for many objects. This information is important in order to define observation programmes on selected targets (feasibility, observatory selection, exposure time definition). As an example,  the Very Large Array Sky Survey (VLASS) is expected to provide  significant coverage of the northern sky with a typical flux density sensitivity of about 0.1\,mJy at 3\,GHz\footnote{https://science.nrao.edu/science/surveys/vlass}. This sensitivity limit is compatible with the emission level of several PACWBs, but is not high enough to detect fainter sources (including potentially many sources not detected in previous observation campaigns). However, upper limits for these sources will be helpful in order to organize potential dedicated observation programmes. Finally, the importance of other surveys such as the Multi-Array Galactic Plane Imaging Survey (MAGPIS; \citealt{magpis}) should not be overlooked. It revealed quite high flux densities (several mJy) at a wavelength of 20 cm for two sources, recently identified to be colliding-wind binaries by \citet{anderson2014}. This does not constitute  compelling evidence for non-thermal radiation, but it appears a priori not compatible with the expected thermal emission from massive star winds. A high brightness temperature, especially at longer wavelengths, is indeed frequently indicative of synchrotron radiation in massive star systems. The information conveyed by such surveys is thus highly valuable for  the identification of new PACWBs.\\

In summary, the favoured observation strategies should consider first a rough estimate of the expected flux densities, either on the basis of existing surveys or through the approach based on Fig.\,\ref{tbd} (the so-called truncated banana diagram). Depending on the radio observatory selected, these numbers should help to define required integration times. Repeated observations are then recommended, ideally at more than one frequency. Finally, for longer period systems, VLBI observations should be considered in order to spatially resolve  the synchrotron emission region. In the case of non-detections, upper limits would be important for investigations with future, more sensitive observatories. This strategy should be applied to many known colliding-wind binaries, selected notably on the basis of multiplicity studies such as those published  by \citet{sana2014} or \citet{vlti2016}, and references therein.

\subsubsection{High-energy observations}\label{high}
When relativistic particles are present in a given astrophysical environment, it is legitimate to envisage that non-thermal high-energy radiative processes are also at work. The most obvious mechanism for electrons is inverse Compton scattering of photospheric photons by the same relativistic electron population involved in the synchrotron process. The upper limits on the Lorentz factors are of the order of 10$^4$--10$^5$ in long-period binaries \citep[e.g.][]{debeckerreview}. The IC photon energy is proportional to the square of the Lorentz factor of the relativistic electrons, and proportional to the incident photon energy as long as the process occurs within the Thomson regime \citep{BG}. Given the typical energies involved for both electrons and photons, Klein-Nishina quantum effects can be neglected in most cases. The typical energy of high-energy photons produced by IC scattering can thus reach values of a few GeV. This process is certainly able to produce non-thermal hard X-rays as demonstrated in the case of $\eta$\,Car \citep{viottietacar,leyderetacar,sekiguchietacar}, and potentially contribute in $\gamma$ rays. The soft X-ray domain (below about 10 keV) is dominated by the thermal emission from the individual stellar winds and from the shocked plasma of the colliding winds. The dominance of thermal processes in the soft X-ray range is a motivation to explore hard X-rays, i.e. above 10\,keV \citep[see e.g.][]{debeckerreview}. This constitutes a key point for observation strategies in the high-energy domain. However, the expected non-thermal component in hard X-rays is certainly too weak to be detected by facilities such as INTEGRAL, as emphasized by \citet{debeckerintegral}. The latter study revealed upper limits in the 20--100 keV range of the order of 10$^{-11}$\,erg\,cm$^{-2}$\,s$^{-1}$. On the basis of models presented by \citet{PD140art} it is  estimated that an improvement in sensitivity  of at least one order of magnitude  in that range is necessary to detect some PACWBs, and certainly one additional order of magnitude is needed to derive spectral properties. The NuStar satellite \citep{nustar} is certainly a highly valuable tool in the search for non-thermal X-ray emission from massive binaries above 10 keV. For instance, the recent modelling of the non-thermal emission from the long-period PACWB HD93129A, depending on various assumptions, predicts a significant detection with NuStar \citep{delpalacio2016}. In this context, future hard X-ray observatories, with even more sensitive instruments, are expected to unveil non-thermal emission components in the X-ray spectrum of massive binaries,  whether they are known to be non-thermal radio emitters or not.\\

The detection of gamma rays from PACWBs remains mostly elusive. Gamma rays are surely produced in all systems with detected synchrotron emission: the electron population traced by the non-thermal radio emission is located in a region of high photon density, and IC interactions producing GeV photons are unavoidable. However, the very same photon field, originating in the hot stars,  is highly opaque to GeV gamma rays  \citep[see e.g.][]{Romeroetal2010}. The result of this high optical depth is the development of electromagnetic cascades \citep[see e.g.][]{Bednarek1997}, with the consequent attenuation of the high-energy photons and the increase in secondary photons with energies near the pair-creation threshold ($~\sim 1$ MeV), a region of the spectrum notoriously difficult to  observe. So far, only $\eta$ Carina has been detected in gamma rays (\citealt{Tavani2009};  see also \citealt{Pshirkov2016} for the potential detection of WR11). Other $\gamma$-ray observations of PACWBs only led to upper limits \citep{wernerfermi}. In the future, MeV-dedicated instruments might be very useful in order to identify high-energy PACWBs, provided they reach sensitivities of the order of $\sim$ 10$^{-11}$\,erg\,cm$^{-2}$\,s$^{-1}$ at 10 MeV, as expected for the AstroMeV mission with an exposure of 100\,ks\footnote{http://astromev.in2p3.fr/?q=aboutus/pact}. We must stress, however,  that given the wide range of parameters characterizing PACWBs in general, quite different emission levels are expected depending on the system that is considered. Theoretical predictions are strongly dependent of some parameters, such as the magnetic field that can drain the energy of electrons before they are able to produce a significant amount of gamma rays by IC scattering of stellar photons. Nevertheless, luminosities in the range 10$^{32}$ -- 10$^{34}$\,erg\,s$^{-1}$ can be expected from sources similar to HD\,93129A \citep[see][]{delpalacio2016}.\\

The conditions likely to lead to a detectable non-thermal emission in X-rays or $\gamma$ rays are significantly different from those which favour the detection of synchrotron radio emission. For instance, a lack of detection of synchrotron radio emission due to an enhanced free-free absorption by the wind material at a given orbital phase will not necessarily be accompanied by the suppression of $\gamma$-ray emission at the same epoch. High-energy observations thus constitute  a relevant complementary approach to identifying particle accelerators among massive binaries. The expected improvement in the sensitivity of high-energy observatories over the forthcoming decades will undoubtedly clarify  our view of PACWBs substantially, with promising prospects for the identification of new members in the catalogue. In particular, the Cherenkov Telescope Array (CTA; \citealt{cta}) is expected to reach a sensitivity slightly better than 10$^{-12}$\,erg\,cm$^{-2}$\,s$^{-1}$ at 100 GeV \citep{cta2}. This should be sensitive enough to detect hadronic emission from PACWBs such as HD93129A, as predicted by \citet{delpalacio2016}. The Fermi telescope, on the other hand, being a survey instrument, is far from ideal for the study of gamma-ray emitting PACWBs. Yearly integration of the emission smothers out the potential periods of significant radiation along the orbit when the right opacity conditions are given in specific systems. It is for this reason that low-threshold ground instruments such as CTA or future pointing missions such as AstroMeV are better options for  unveiling the subclass of gamma-ray colliding wind massive binaries, of which $\eta$\,Car is the prototype.\\

\section{Summary and conclusions}\label{concl}

On the basis of several criteria, we qualitatively explored the issue of the fraction of particle accelerators (the so-called PACWBs) among colliding-wind binaries. The main ideas arising from our critical discussion are summarized as follows:
\begin{enumerate}
\item[-] Future identifications of PACWBs are expected for massive binary system harbouring O and WR stars. However, a lower priority is  attributed for late  main-sequence O-stars because their winds are probably not powerful enough to drive non-thermal processes up to a level reachable by most current observatories.
\item[-] To date, the most favoured massive binary systems likely to be investigated to identify new PACWBs are systems with periods longer than a few weeks, with no clear anticipated upper boundary.
\item[-] Even though thermal X-ray emission from colliding-winds may be viewed as a good indicator of energy transfer in the wind-wind interaction, faint X-ray emitting colliding-winds may be significant non-thermal emitters.  
\item[-] Investigations should not be restricted to stars known to produce magnetic fields detectable by current observatories. Surface magnetic fields below the gauss level may be enough to feed synchrotron radio emission in the wind-wind interaction region.
\end{enumerate}

On the basis of these considerations, there is no clear reason to consider particle acceleration in massive binaries as an anomaly, or even as a rare phenomenon; no clear criterion has yet been identified that is likely to drastically restrict the actual fraction of PACWBs among massive binaries. Assuming a population of about 10$^5$ massive binaries (i.e. a conservative value assuming a massive star population fraction of 10$^{-6}$ of the Milky Way stellar population, including at least 50\% binaries), with a significant fraction of binaries with periods larger than a few weeks, one cannot exclude that the Galactic PACWB population may exceed 10$^4$ members. Such a value should by no means constitute a solid prediction, but rather a reasonable guess based on current knowledge on massive star populations. The actual census of PACWBs thus either suffers from a severe observational bias or suggests that a still unidentified factor influences the non-thermal activity of these objects. Many observations are thus required to improve  the inventory of particle accelerators significantly and to achieve a much better view of the fraction of PACWBs among massive binaries. In this context, radio observations are expected to be the main providers of relevant information, even though highly significant prospects are expected from high-energy investigations. The intensive search for new members in the catalogue constitutes an important requirement to estimate the contribution of the population of colliding-wind binaries to the production of soft Galactic cosmic rays.

\begin{acknowledgements}
M. De Becker acknowledges the financial support from ULg through a `Cr\'edit classique (DARA)'. G.~E. Romero acknowledges support by the Spanish Ministerio de Economía y Competitividad (MINECO/FEDER, UE) under grant AYA2016-76012-C3-1-P and CONICET grant PIP 0338. The authors thank the anonymous referee for comments that allowed us to clarify and significantly improve the paper. The SIMBAD database was used for the bibliography.
\end{acknowledgements}

\bibliographystyle{aa}




\end{document}